# Plasmon-Enhanced Interface Charge Transfer Exciton of 2D In-Plane Lateral and van Der Waals MoS$_2$/WS$_2$ Heterostructures

*Xijiao Mu, Mengtao Sun*[*]

School of Mathematics and Physics, Center for Green Innovation, Beijing Key Laboratory for Magneto-Photoelectrical Composite and Interface Science, University of Science and Technology Beijing, Beijing 100083, People's Republic of China.

No. 30 Xueyuan Road, Beijing 100083, P. R. China

Email: mengtaosun@ustb.edu.cn (M. T. Sun)

**Abstract**: The multi-scale computational method of combining first-principles calculation and finite element electromagnetic simulations is used to study plasmon-enhanced interlayer charge transfer (ICT) exciton of 2D lateral and van der Waals MoS$_2$/WS$_2$ heterostructure with 2H phase. The weak ICT excitons are observed in the 2H lateral and van der Waals MoS$_2$/WS$_2$ heterostructures. Theoretical results reveal the physical principle of plexciton resulting from strong coupling between plasmonic and ICT exciton. The weak CT exciton can be strongly enhanced by metal plasmon, can provide a new way to observe the weak CT exciton. Our results can promote deeper understanding of the plexciton resulting from strong coupling interaction between plasmon and exciton of lateral and van der Waals heterostructures.

## 1.Introduction

Two-dimensional layered transition metal dichalcogenides (TMDs) semiconductors are one of the best candidates for next-generation optoelectronics, electronics and valleytronic devices, [1-10] due to their direct energy bandgap and the non-centrosymmetric lattice structure.[10-12] TMDs open up a new avenue for van der Waals (vdW) and/or lateral heterostructures with great design flexibility.[13-17] It is challenging for the fabrication of 2D heterostructures with clean and sharp interfaces, which is essential for preserving optoelectronic properties driven by the intralayer or interlayer coupling.

The van der Waals heterostructures, built up from monolayer TMDs, are held together by vdW forces in the absence of dangling bonds, which can offer atomically regulated interfaces and thereby provide sharp band edges. Charge transfer excitons at interfaces of vdW heterostructures can give rise to bound electron−hole pairs across the interface such as charge transfer (CT) or interlayer excitons.[18] The interface excitons have been successfully observed in TMDs vertical heterostructures,[18-22] which can drive an efficient transfer of photogenerated charge carriers, such that the electrons accumulated in one layer and the holes in the other layer, forming indirect or interlayer excitons with longer electron−hole recombination lifetime.[19,22-24] The exciton decay dynamics is dominated by interlayer charge transfer (CT) processes, while fast interlayer energy transfer (ET) in $MoS_2$/$WS_2$ heterostructures was also observed experimentally using photoluminescence excitation spectroscopy.[25] When used as active layers in optoelectronic devices, they can potentially overcome some limitations of individual TMDs layers and possess different electronic structures and optical properties from TMDs heterostructures. The vdW heterostructures cannot be precisely controlled, and the interface between layers hampered by defects and contamination in the processes.[26,27]

The in-plane lateral heterostructures of monolayer TMDs with atomically sharp interfaces without interfacial contamination in the heterostructures are highly demanded, because of a single crystalline signature of the TMDs atomic layers and a unique 1D heterointerface with no discernible height difference.[28-34] Because of the difficulties in controlling the defects and contamination of 2D uniaxial crystals, the fabrication of 2D heterostructures with pure and sharp interfaces is still challenging to maintain the photoelectric performance driven by interlayer or intralayer coupling.[35] Therefore, it is necessary to investigate the optical properties theoretically. The atomically sharp interface of lateral heterostructure offers an interesting platform for the study of fundamental materials science of essential building block of 2D electronics and optoelectronic components in the future. Optoelectronic properties of lateral heterostructures are determined by the energy band structure, doping, and defects of both materials near the boundary.[29,30] The type II (staggered) band alignment is anticipated at the heterostructure, which may lead to the formation of excitons by electrons and holes localized on opposite sides of the 1D heterointerface, enabling p–n junctions and efficient optoelectronics.[19]

Whether it is a vdW heterostructure or a lateral heterostructure, the heterostructure of TMDs has extraordinary application potential in the field of photocatalysis and optoelectronic devices. The two-dimensional materials of most TMDs have both metal and semiconductor phases. This two-phase nature is inherently determined by the heterostructure being a metal-semiconductor contact. When the heterostructure is exposed to a certain amount of light, the metal phase produces surface plasmons, and excitons are present in the semiconductor phase. Therefore, it is urgent to study the coupling properties of plasmonic and exciton in TMDs heterostructure materials. Coherently coupled plasmons and excitons can result in new optical excitations: plexcitons, which is polaritonic modes, due to the strong coupling of these two oscillator systems. The strong coupling between the surface plasmon and the exciton will also cause the appearance of Rabi-splitting, that is, the very close absorption peaks will split into two closer resonance peaks. This strong coupling causes changes in the electromagnetic field mode and exciton wave function.[36-38] Physical mechanism and applications of strong and weak coupling have been extensively investigated.[39-43] In general, the spectral region of strong coupling where the imaginary part of the dielectric function is large positive, and the real part large negative is taken as an evidence for the plasmon.[44~47] While, the interaction between plasmon and excitonic plasmon interaction is not clear, especially principle of coupling interaction between plasmon and CT excitons of lateral and van der Waals heterostructures, since the absorption intensity of CT exciton itself is usually too weak to be observed experimentally. In addition to strong coupling, there is a relatively weak coupling, also known as the Purcell effect. This weak coupling effect can also significantly increase the exciton density, but it will not change the electromagnetic field mode and exciton wave function. Therefore, for the enhancement of the charge transfer exciton, the weak coupling mode can keep the charge transfer behavior unchanged.[48,49] It is very important to reveal the physical mechanism that the metal plasmon enhances CT excitons, and make the CT excitonic peak is strong enough to be clearly observed experimentally.

In this works, plasmon-enhanced nanophotonic properties of van der Waals and lateral 2H phase $MoS_2/WS_2$ heterostructure with investigated in detail theoretically. The optical parameters calculated by the density functional theory (DFT) method can effectively describe the structural changes of materials at the atomic level. The combination of this optical parameter and finite element method (FEM) can analyze the interaction mechanism of plasmons and excitons on the subatomic scale. This method effectively resolves the problem that the traditional electromagnetic field theory cannot describe the quantum effect well in the subatomic field.

**2. Method**

In this letter, the first-principles calculation, which calculate and analysis without experimental data, was used as a major computational technique. The calculations were performed by the Vienna Ab-initio Simulation Package (VASP).[50] With The 11.0×11.0×1.0 Γ-centered k-points in k-space, the geometry optimization, total energies, forces, and energy profiles are calculated using density function theory (DFT)[51] combine spin-orbit coupling algorithm. The project augmented planewave (PAW) method [52] and plane wave basis set are used in VASP, and the Perdew-Burke-Ernzerhof (PBE)[53] functional which belongs to generalized gradient approximation (GGA)[54] approach was adopted for exchange–correlation energy. To accurately consider the nonbond interaction (mainly dispersion interaction) between layers, we used the DFT-D3.[55] An energy cut-off of 500 eV is used for plane-wave expansion of the PAW. A vacuum space of 18 Å in the Z-direction is introduced to eliminate the interaction between layers. This model is an effective single layer approximation method. The geometry was optimized until the Hellmann-Feynman force were less than 0.0005 eV/Å and the total energy is converged within $10^{-8}$ eV. The Bethe-Salpeter equation (BSE)[56] belong to many-body perturbation theory is used in optical property calculations to obtain a dielectric function with higher precision. The BSE method can effectively consider the exciton effect inside the material, which can further improve the accuracy of the calculation absorption. The exciton excitation is derived from the GW approximation,[57] so BSE needs to rely on the GW approximation of the wave function.[58] For lateral heterostructures, we built an 8×4×1 supercell. In this case, 1.0×1.0×1.0 k-point sampling is used. Full optimization of the atomic geometry was performed until all components of the residual forces were less than 0.05 eV/Å and the total energy is converged within $10^{-6}$ eV. On the other hand, in this work, we use the DFT method to calculate the anisotropic dielectric function as a parameter for finite element analysis. Frequency domain analysis was performed using the wave optics module in the Comsol Multiphysics 5.4a software. During the modeling process, the Ag disk is placed in the center of the cuboid, and the rectangular body is surrounded by periodic boundary conditions and combined with a perfect electrical conductor. The top and bottom of the cuboid are two ports. The electromagnetic field equation uses the optical equations calculated by the anisotropic equations in combination with the DFT method.

## 3. Results and discussions

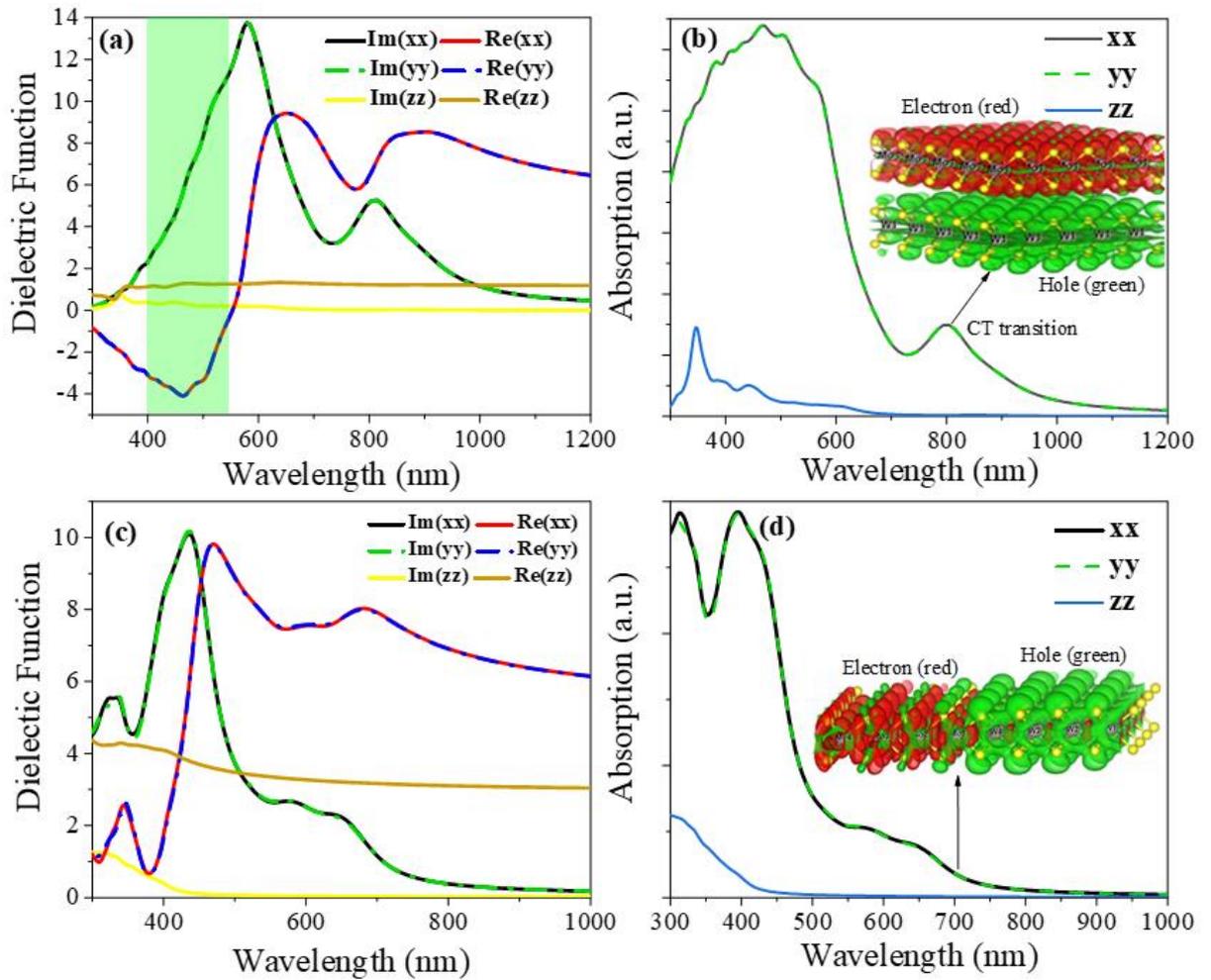

**Figure 1.** The anisotropy dielectric function (a) and absorption spectrum (b) of vdW heterostructure of $MoS_2/WS_2$. The anisotropy dielectric function (a) and absorption spectrum (b) of lateral heterostructure that constituted by $MoS_2/WS_2$.

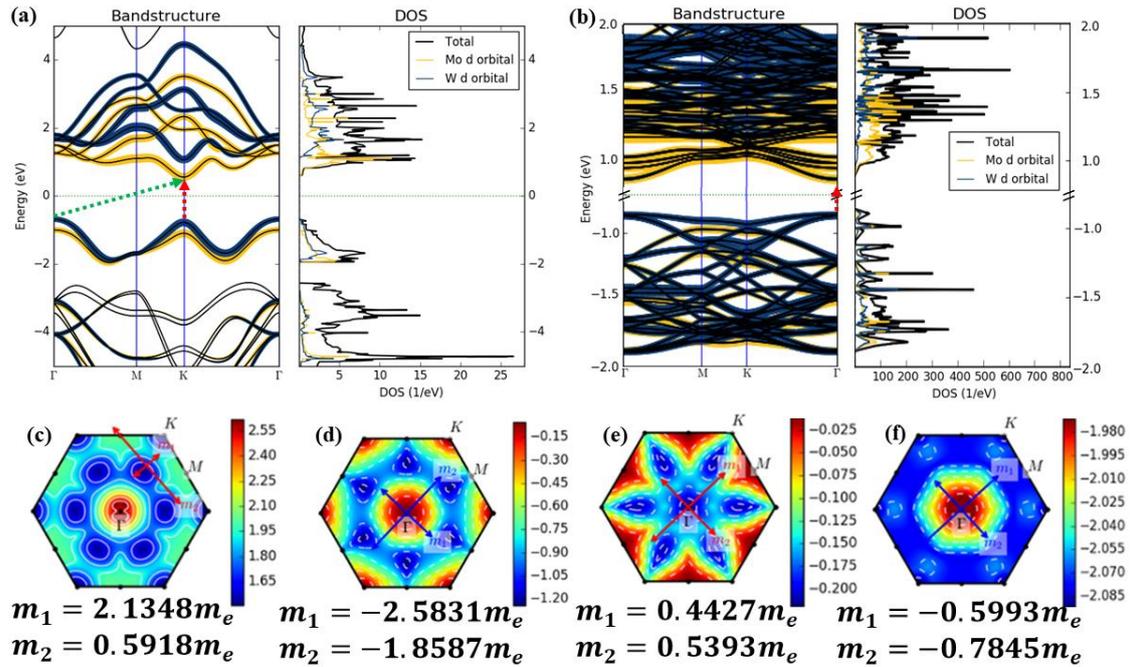

**Figure 2.** The electronic structure (energy band and density of states) of vdW (a) and lateral (b) 2H phase heterostructure that constituted by $MoS_2/WS_2$. The green and red arrow stand for the indirect and direct band gap transition, respectively. The color map of first Brillouin zone electron and hole effective mass of vdW (c, d) and lateral (e, f) heterostructure.

Before studying excitons and plasmon, we must calculate the exact optical properties. The complex dielectric function of the material can well reflect the system and light interaction. When the complex dielectric function of the system is less than zero and the imaginary part is large, it exhibits certain metallic properties. Fig. 1(a) is the complex dielectric function of vdW heterostructure. It is found that there are large negative real part and imaginary part of dielectric function around 465 nm for 2H van der Waals heterostructure, so there are strong light-matter interactions, which demonstrates strong interaction of excitonic and plasmonic. Also, Fig. 1(b) shows strong optical absorption around 465 nm for 2H phase van der Waals heterostructure. Due to the van der Waals interaction for the heterostructure, the strong CT excitons can be well observed around 800 nm for 2H phase heterostructure. The insert isosurface represent the electron-hole density, which is defined by:

$$n_e(\mathbf{r}) - n_g(\mathbf{r}) = \sum_{n,k} \psi^*_{n\mathbf{k}e}(\mathbf{r})\psi_{n\mathbf{k}e}(\mathbf{r}) - \sum_{m,k} \psi^*_{m\mathbf{k}g}(\mathbf{r})\psi_{m\mathbf{k}g}(\mathbf{r}) \qquad (1)$$

where the $n_e(\mathbf{r})$ and $n_g(\mathbf{r})$ stand for density of excited (e) and ground state (g), respectrively. The $\psi_{m\mathbf{k}e}(\mathbf{r})$ represent the excited state PAW wavefunction of $m$ orbital at $k$ point. Note the interlayer CT (ICT) peak around 800 nm for 2H heterostructure is pure interlayer excitons, where the electron and hole are distributed $MoS_2$ and $WS_2$, respectively. There can be two transition modes in the projected energy band: indirect band gap transitions with lower energy (green arrow) and direct band gap transitions with higher energy (red arrow), see Fig. 2(a, b). Yellow and blue in the projected energy bands represent Mo and W atoms, respectively. The low-energy transitions near the Fermi level are all transitions between different color bands. So, it must be a charge transfer transition. By combining the difference in density of the two transitions with Eq.1, a strong charge transfer exciton can be revealed. For lateral heterostructure, the anisotropy dielectric function is different from vdW heterostructure, see Fig. 1(c). It is found that there is no large negative real part for 2H phase lateral heterostructures, so there is no strong light-matter interaction, which demonstrates there is no strong interaction between exciton and plasmonic. Fig. 1(d) show strong optical absorption from 300 nm to 500 nm, and weak absorption from 500 to 700 nm for 2H lateral heterostructures. Combining the absorption spectrum and the dielectric function, it can be found that the protrusion of the dielectric function around 700 nm is caused by the intralayer exciton. These CT transfer can be well demonstrated, see the inset in Fig. 1(d). The lowest point of the lateral heterojunction conduction band is in the Brillouin area, which changes from K to Γ. Therefore, the indirect transitions in the lateral heterojunction disappear and are replaced by direct band gap transitions at the Γ point (the red arrow in the Fig. 2(b)). On the other hand, the electron and hole effective mass can determine the SPs frequency. In the surface physical optics, the SPs frequency is defined by:

$$\omega_p = \sqrt{\frac{ne^2}{\varepsilon_0 m^*}}$$

where the $n$, $e$ and $\varepsilon_0$ are the carrier concentration, electron charge in SI units and the vaccum dielectric constant, respectively. The $m^*$ is the electron or hole effective mass. The electron and hole effective mass are positive and negative, respectively. For the 2D materials, the effective is anisotropic and divide to two component. As shown in Fig. 2, in the wdW heterostructure, the effective mass anisotropy of electrons and holes is strong, indicating that the plasmon frequencies in different directions also have strong anisotropy, see Fig. 2(c, d). However, a larger effective mass will generate low-frequency plasmons, which indicates that vdW heterostructures are more likely to generate low-frequency SPs than lateral heterostructures, see Fig. 2(e, f).

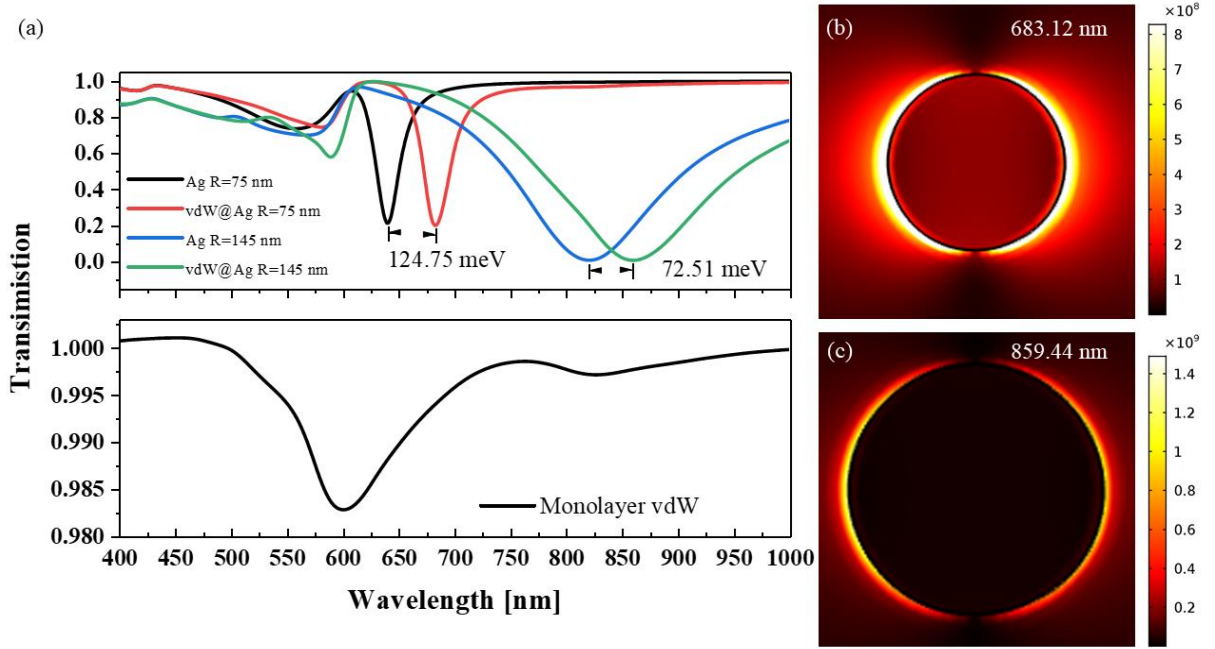

**Figure 3.** Finite Element Analysis of Electromagnetic Field of vdW heterostructure and Ag Discs with the dielectric function that calculated by BSE. (a) Transmission spectra of coupling between different size Ag discs of and vdW heterostructure. The dipole electromagnetic field mode of vdW heterostructure@Ag disc at 683.12 nm (b) and 859.44 nm (c).

According to the previous discussion, there are both surface plasmons and charge transfer excitons in the vdW heterostructure. Therefore, a special size (R=75 nm) is set for the Ag disc so that its SPR peak is as close as possible to the SPR peak of the vdW heterostructure. There is a strong coupling between the two, causing the original SPR peak (the bottom half of Fig 3(a)) to shift by 124.75 meV and a Rabi-splitting, see Fig 3(a). The strong coupling will change the original electromagnetic field mode, making the original dipole mode field wider and the strong field area larger, see Fig 3(b). On the other hand, changing the radius of the Ag disk to 145 nm makes the SPR peak of Ag sufficiently close to the CT exciton absorption peak of the vdW heterostructure. In this case, there is a weak coupling (shift by 72.51 meV) between the surface plasmonic of Ag and the CT exciton of vdW, and it is a Purcell effect. In other words, the wave functions of the exciton and electromagnetic modes of the plasmon are unperturbed. The width of the electromagnetic field mode is small.

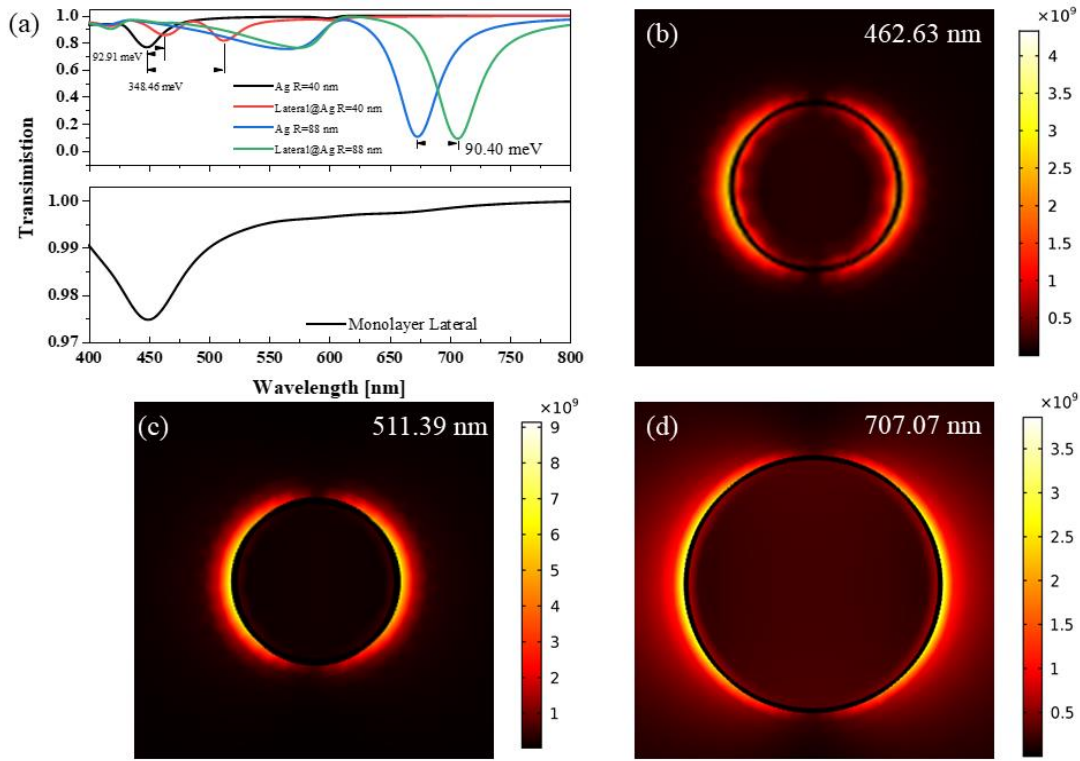

**Figure 4** Finite Element Analysis of Electromagnetic Field of lateral heterostructure and Ag Discs with the dielectric function that calculated by ab initio. (a) Transmission spectra of coupling between different discs of and lateral heterostructure; The dipole electromagnetic field mode of lateral heterostructure@Ag disc at 426.63 nm (b), 511.39 nm (c) and 707.07 nm (d).

For the lateral heterostructure, because there is no surface plasmon, the surface plasmon of the Ag disc is simply coupled with the exciton in the heterostructure. Firstly, when the radius of the Ag disk is 40 nm, its transmission spectrum is as close as possible to the position of the strong absorption peak of the lateral heterostructure.

A Rabi-splitting[36,38] appears in the transmission spectrum in Fig. 4(a) and the SPR peaks shift by 92.91 meV and 348.46 meV, respectively. It splits into two strong peaks, and the electromagnetic field modes at this time are shown in Fig. 4(b) and (c), respectively. The electromagnetic fields of these two electric field modes have also changed, from a dipole mode to a superposition of a dipole mode and a multipolar mode. Change the radius of the Ag disk to 88nm to make it weakly coupled, which is the Purcell effect at 707.07 nm and the SPR peak shift by 90.4 meV. The electric field mode is shown in Fig. 4(d). The electric mode field is not changed, and is still a dipole mode.

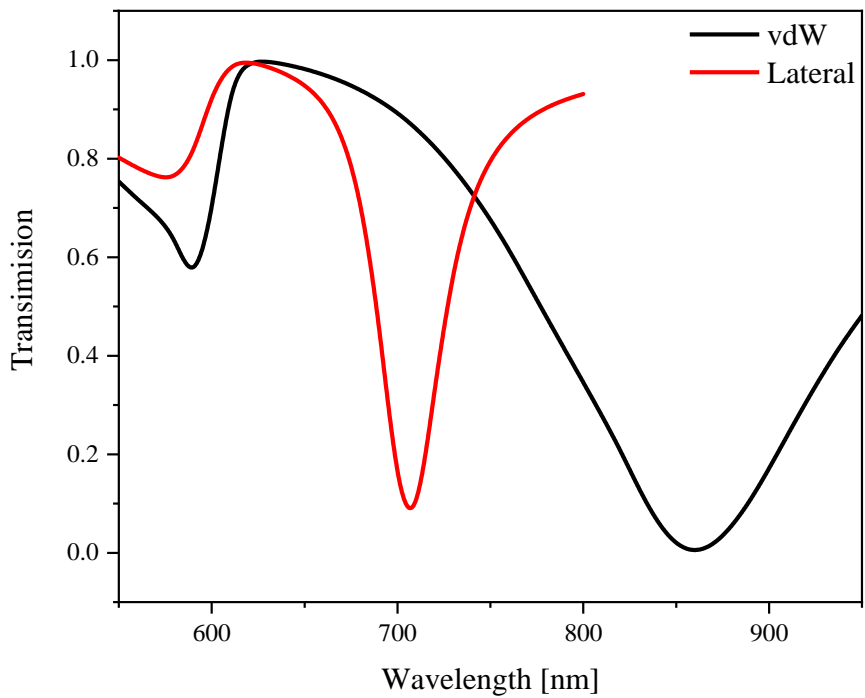

**Figure 5** Relative relationship of transmission spectra in different material configurations

According to the discussion in the previous two sections, in the long-wavelength region, a weak coupling occurs between the surface plasmon of the Ag disk and the charge transfer exciton in the heterostructure when they interact. This weak coupling belongs to the Purcell effect. This effect does not change the mode of the surface plasmon electromagnetic field and the exciton wave function, but also can obtain a good field strength. Therefore, this method can be used to enhance the charge transfer exciton without changing the charge transfer method. In other words, only increasing the intensity of charge transfer does not change the way of charge transfer.

The finite element analysis of classical electromagnetism has practical difficulties in dealing with the interlayer interactions of two-dimensional materials. In this work, we used density functional theory (DFT) and many-body perturbation theory to calculate the optical properties of two-dimensional materials, especially the dielectric function. Therefore, by analyzing this effect by transmission spectroscopy, we find that the wavelength of the transmission spectrum peaks is red shifted in the presence of charge transfer, see Fig. 5. According to the conclusion of the CDD calculated by the previous DFT, the vdW heterostructure has the most complete charge transfer, so the red shift is the most. Surface plasmons can significantly enhance charge transfer because the hot electrons have a higher transport capacity, and cluster excitation can promote charge transfer at the interface.[59] The lateral heterogeneous structure is second, so the redshift is relatively weak. The peak wavelength of the transmission spectrum satisfies the following relationship:

$$\lambda_{vdW} > \lambda_{Lateral} \qquad (2)$$

This regular can be used to determine the existence of interlayer and intralayer CT excitons, as well as the type of heterostructure in finite element analysis.

**Conclusions**

The strong and weak interaction with Rabi-splitting, Purcell effect and plexciton between metal plasmon and $MoS_2/WS_2$ heterostructure CT exciton are reported in this letter. The strong interaction like plexciton and Rabi-splitting exist in the high energy region. Although this interaction enhances the electromagnetic field, it will change the exciton wave function and electromagnetic field mode. However, the weak coupling, Purcell effect, is a better method to enhanced CT exciton. This method can enhance the charge transfer exciton density without changing the electromagnetic field mode and exciton wave function. Weak coupling positions of different types of heterostructures are different. The wavelength of the vdW heterojunction is longer. This rule can be used to judge the type of heterostructure and the type of charge transfer.

**Corresponding Author**

*E-mail: mengtaosun@ustb.edu.cn.

**ORCID**

Xijiao Mu: 0000-0002-7991-0669

Mengtao Sun: 0000-0002-8153-2679

**Acknowledgments**

This work was supported by the National Natural Science Foundation of China (91436102, 11374353 and 11874084), the National Basic Research Program of China (Grant number 2016YFA0200802), and the fundamental Research Funds for the Central Universities.

**References**

[1]. Mak, K. F.; Lee, C.; Hone, J.; Shan, J.; Heinz, T. F. Atomically thin MoS$_2$: A new direct-gap semiconductor. *Phys. Rev. Lett.* **2010**, 105, 136805-136808.

[2]. Radisavljevic, B.; Radenovic, A.; Brivio, J.; Giacometti, V.; Kis, A. Single-layer MoS$_2$ transistors. *Nat. Nanotechnol.* **2011**, 6, 147–150.

[3]. Wang, Q. H.; Kalantar-Zadeh, K.; Kis, A.; Coleman, J. N.; Strano, M. S. *Nat. Mater.* **2012**, 7, 699−712.

[4]. Zeng, H.; Dai, J.; Yao, W.; Xiao, D.; Cui X Valley polarization in MoS$_2$ monolayers by optical pumping. *Nat. Nanotechnol.* **2012**, 7, 490–493.

[5]. Jones, A. M.; Hongyi, Y.; Nirmal, J. G.; San, F, W.; Grany, A. Optical generation of excitonic valley coherence in monolayer WSe$_2$. *Nat. Nanotechnol*. **2013**, 8, 634–638.

[6]. Britnell, L.; Ribeiro, R. M.; Eckmann, A.; Jalil, R.; Belle, B. D.; Mishchenko, A.; Kim, Y. J.; Gorbachev, R.V.; Georgiou, T.; Morozov, S. V. Strong Light-Matter Interactions in Heterostructures of Atomically Thin Films. *Science*. **2013**, 340, 1311−1314.

[7]. Zhang, Y. J.; Oka, T.; Suzuki, R.; Ye, J. T.; Iwasa, Y. Electrically Switchable Chiral Light-Emitting Transistor. *Science*. **2014**, 344, 725−728.

[8]. Manish, C.; Hyenon, S. S.; Goki, E.; Lain, J. L.; Kian, P.; Hua, Z. The chemistry of two-dimensional layered transition metal dichalcogenide nanosheets. *Nat. Chem.* **2013**, 5, 263-275.

[9]. Xi, D. D.; Chen, W.; Anlian, P.; Ruqin, Y.; Xiang, F. D. Two-dimensional transition metal dichalcogenides as atomically thin semiconductors: opportunities and challenges. *Chem. Soc. Rev.* **2015**, 44, 8859-8876.

[10]. Ji, Q.; Zhang, Y.; Liu, Z. Chemical vapour deposition of group-VIB metal dichalcogenide monolayers: engineered substrates from amorphous to single crystalline. *Chem. Soc. Rev.* **2015**, 44, 2587-2602.

[11]. Bonaccorso, F.; Sun, Z.; Hasan, T.; Ferrari, A. Graphene photonics and optoelectronics. *Nat. Photonics.* **2010**, 4, 611−622.

[12]. Avouris, P. Graphene: electronic and photonic properties and devices. *Nano Lett.* **2010**, 10, 4285−4294.

[13]. Geim, A. K.; Grigorieva, I. V. Van der Waals heterostructures. *Nature*, **2013**, 499, 419–425.

[14]. Yu, W. J.; Zheng, L.; Yu, C.; Hailong, Z.; Yang, W.; Xiang, F. D. Vertically stacked multi-heterostructures of layered materials for logic transistors and complementary inverters. *Nat. Mater.* **2013**, 12, 246–252.

[15]. Georgiou, T.; Rashid, J.; Branson, D. B.; Liam, B.; Roman, V. G.; Sergey, V. M.; Yong, J. K.; Laurence, E.; Artem, M. L. Vertical field-effect transistor based on graphene-WS$_2$ heterostructures for flexible and transparent electronics. *Nat. Nanotechnol.* **2013**, 8, 100–103.

[16]. Yu, W. J.; Yuan, L.; Hailong, Z.; Anxiang, Y.; Zheng, L.; Huang. Y.; Xiang. F. D. Highly efficient gate-tunable photocurrent generation in vertical heterostructures of layered materials. *Nat. Nanotechnol.* **2013**, 8, 952–958.

[17]. Gupta, A.; Sakthivel, T.; Seal, S. Recent development in 2D materials beyond graphene. *Prog. Mater. Sci.* **2015**, 73, 44–126.

[18]. Anupum, P.; Zafer, M.; Darshana, W.; Hui, C.; Roger, K. L.; Cengiz, O.;

Sefaattin, T. Fundamentals of lateral and vertical heterostructures of atomically thin materials. *Nanoscale*, *2016*, 8, 3870-3887.

[19]. Li, M.Y.; Chen C. H.; Shi, Y.; Li, L. J. Heterostructures based on two-dimensional layered materials and their potential applications. *Mater. Today* *2016*, 19, 322-335.

[20]. Jun, K.; Jing, L.; Shu, L.; Jian, B.X.; Wang, W. W. Electronic Structural Moiré Pattern Effects on $MoS_2$/$MoSe_2$ 2D Heterostructures. *Nano Lett.* *2013*, 13, 5485−5490.

[21]. Xiao, Y. Z.; Nicholas, R.M.; Zi, Z. G. Haiming, Z.; Kristopher, W. W.; Cory, A. N. Charge Transfer Excitons at van der Waals Interfaces. *J. Am. Chem. Soc.* *2015*, 137, 8313−8320.

[22]. Rivera, P.; John, R. S.; Sanfeng. W.; Grant. A.; Philip, K.; Kyle, S.; Genevieve, G.; Wang, Y.; Xu, X.D. Observation of long-lived interlayer excitons in monolayer $MoSe_2$–$WSe_2$ heterostructures. *Nat. Commun.* *2015*, 6, 6242-6247.

[23]. Bastian, M.; Alexander, S.; Borja, P.; Julian, K.; Frank, J.; Alexander, H.; Ursula, W. Long-Lived Direct and Indirect Interlayer Excitons in van der Waals Heterostructures. *Nano Lett.* *2017*, 17, 5229−5237.

[24]. Ross, J. S.; Rivera, P.; Schaibley, J.; Lee-Wong, E.; Yu, H.; Taniguchi, T.; Watanabe, K.; Yan, J.; Mandrus, D.; Cobden, D.; Yao, W.; Xu, X. Interlayer Exciton Optoelectronics in a 2D Heterostructure p–n Junction. *Nano Lett.* *2017*, 17, 638−643.

[25]. Schaibley, J. R.; Rivera, P.; Yu, H.; Seyler, K. L.; Yan, J.; Mandrus, D. G.; Taniguchi, T.; Watanabe, K.; Yao, W.; Xu, X. Directional interlayer spin-valley transfer in two-dimensional heterostructures. *Nat. Commun.* *2016*, 7, 13747-13752.

[26]. Nagler, P.; Plechinger, G.; Ballottin, M. V.; Mitioglu, A.; Meier, S.; Paradiso, N.; Strunk, C.; Chernikov, A.; Christianen, P. C. M.; Schüller, C.; Korn, T. Interlayer exciton dynamics in a dichalcogenide monolayer heterostructure. *2D Mater.* ***2017***, 4, 025112-025126.

[27]. Rivera, P.; Seyler, K. L.; Yu, H.; Schaibley, J.; Yan, J.; Mandrus, D. G.; Yao, W.; Xu, X. Valley-polarized exciton dynamics in a 2D semiconductor heterostructure. *Science*, ***2016***, 351, 688−691.

[28]. Hong, X.; Kim, J.; Shi, S. F.; Zhang, Y.; Jin, C.; Sun, Y.; Tongay, S.; Wu, J.; Zhang, Y.; Wang, F. Ultrafast charge transfer in atomically thin MoS2/WS2 heterostructures. *Nat. Nanotechnol.* ***2014***, 9, 682−686.

[29]. Dai, C. K.; Alexandra, C. Ivan, V.; Francenso, G.; Yuhei, M.; Shinichiro, M.; Goki, E. Evidence for Fast Interlayer Energy Transfer in MoSe$_2$/WS$_2$ Heterostructures. *Nano Lett.* ***2016***, 16, 4087−4093.

[30]. Haigh,S.J.;Gholina,A.;Romani,S.;Brtinell,L.;Elias,D.C.;Ponomarenko,L.A.; Geim,A.K.;Grobachev,R. Cross-sectional imaging of individual layers and buried interfaces of graphene-based heterostructures and super lattices. *Nat. Mater.* ***2012***, 11, 764–767.

[31]. Yang, W.; Guo. R. C.; Zhi, W. S.; Cheng, C. L. Epitaxial growth of single-domain graphene on hexagonal boron nitride. *Na. Mater.* ***2013***, 12, 792–797.

[32]. Zhang, Q.; Lin, C.; Tseng, Y.; Huang, K.; Lee, Y. Synthesis of Lateral Heterostructures of Semiconducting Atomic Layers. *Nano Lett.* ***2015***, 15, 410−415.

[33]. Shi, Y.; Li, H.; Li, L.J. Recent advances in controlled synthesis of two-dimensional transition metal dichalcogenides via vapour deposition

techniques. *Chem. Soc. Rev.* **2015**, 44, 2744-2756.

[34]. Huang, C.; Wu, S.; Sanchez, A. M.; Peters, J. J. P.; Beanland, R.; Ross, J. S.; Rivera, P.; W, Yao. D. H.; Cobden.; Xu, X.; Lateral heterostructures within monolayer MoSe$_2$–WSe$_2$ semiconductors. *Nat. Mater.* **2014**, 13, 1096-1101.

[35]. Gong, Y.; Lin, J.; Wang, X.; Shi, G.; Lei, S.; Lin, Z.; Zou, X. Vertical and in-plane heterostructures from WS$_2$/MoS$_2$ monolayers. *Nat. Mater.* **2014**, 13, 1135-1142.

[36]. Amendola, V.; Pilot, R.; Frasconi, M.; Maragò, O. M.; Iatì M. A. Surface plasmon resonance in gold nanoparticles: a review. *J. Phys. Condens. Matter.*, **2017**, 29, 203002-20350.

[37]. Tame, M. S.; McEnery, K. R.; Özdemir, Ş. K.; Lee, J.; Maier, S. A.; Kim, M. S. Quantum plasmonics. *Nat. Phys*. **2013**, 9, 329–340

[38]. Khurgin, J. B. How to deal with the loss in plasmonics and metamaterials. *Nat. Nanotechnol*. **2015**, 10, 2–6.

[39]. Duan, X.; Wang, C.; Shaw, J. C.; Cheng, R.; Chen, Y.; Li, H.; Wu, X. Lateral epitaxial growth of two-dimensional layered semiconductor heterostructures. *Nat. Nanotechnol.* **2014**, 9, 1024-1030.

[40]. Li, M. Y.; Shi, Y.; Cheng, C. C.; Lu, L.S.; Lin, Y. C.; Tang, H. L.; Tsai, M. L. Epitaxial growth of a monolayer WSe$_2$–MoS$_2$ lateral p-n junction with an atomically sharp interface. *Science.* **2015**, 349, 524-528.

[41]. Tsai, M.; Li, M.; Retamal, J.; Lam, K.; Lin, Y.; Suenaga.; Chen, L.; Liang, G.; Li, L.; He, J. Single Atomically Sharp Lateral Monolayer p-n Heterostructure Solar Cells with Extraordinarily High Power Conversion Efficiency. *Adv. Mater.* **2017**, 29, 1701168-1701174.

[42]. Chen, K.; Wan, X.; Xie, W.; Wen, J.; Kang, Z.; Zeng, X.; Chen, H.; Xu, J. Lateral Built-In Potential of Monolayer $MoS_2$–$WS_2$ In-Plane Heterostructures by a Shortcut Growth Strategy, *Adv. Mater.* **2015**, 17, 6431–6437.

[43]. Yoo, Y.; Degregorio, Z. P.; Johns, J. E. Seed Crystal Homogeneity Controls Lateral and Vertical Heteroepitaxy of Monolayer $MoS_2$ and $WS_2$, *J. Am. Chem. Soc.* **2015**, 137, 14281−14287.

[44]. Du, G. X., Mori, T., Suzuki, M., Saito, S., Fukuda, H., Takahashi, M. (2010). Evidence of localized surface plasmon enhanced magneto-optical effect in nanodisk array. *Appl. Phys. Lett.*, **2010**, 96, 081915.

[45]. González‐Díaz, J. B.; García‐Martín, A.; García‐Martín, J. M.; Cebollada, A.; Armelles, G.; Sepúlveda, B.; Alaverdyan, Y.; Käll, M., Plasmonic Au/Co/Au Nanosandwiches with Enhanced Magneto‐optical Activity, *Small*, **2008**, 4, 202-205.

[46]. Novotny, L.; Hecht, B., Principles of nano-optics. Cambridge university press: 2012.

[47]. Du, G. X., Mori, T., Suzuki, M., Saito, S., Fukuda, H., Takahashi, M.. Evidence of localized surface plasmon enhanced magneto-optical effect in nanodisk array. *Appl. Phys. Lett.*, **2010**, 96, 081915

[48]. Hümmer, T.; García-Vidal, F.; Martín-Moreno, L.; Zueco, D., Weak and strong coupling regimes in plasmonic QED, Phys Rev B, 2013, 87, 115419.

[49]. Purcell, E. M.; Torrey, H. C.; Pound, R. V. Resonance absorption by nuclear magnetic moments in a solid. *Phys. Rev.* **1946**, 69, 37–38.

[50]. Pelton, M. Modified spontaneous emission in nanophotonic structures. *Nat. Photonics*. **2015**, 9, 427–435.

[51]. Kresse, G.; Joubert, D. From ultrasoft pseudopotentials to the projector augmented-wave method. *Phys. Rev. B.* **1999**, 59, 1758-1775.

[52]. Kohn, W.; Sham, L. J. Self-consistent equations including exchange and correlation effects. *Phys. Rev.* **1965**, 140,1133-1138.

[53]. Vanderbilt, D.; Soft self-consistent pseudopotentials in a generalized eigenvalue formalism. *Phys. Rev. B.* **1990**, 41, 7892-7895.

[54]. Perdew, J. P.; Burke, K.; Ernzerhof, M. Generalized gradient approximation made simple. *Phys. Rev. Lett.* **1996**, 77, 3865-3868.

[55]. Vanderbilt, D.; Soft self-consistent pseudopotentials in a generalized eigenvalue formalism. *Phys. Rev. B.* **1990**, 41, 7892-7895.

[56]. Grimme. S. Semiempirical GGA‐type density functional constructed with a long-range dispersion correction. *J. Comput. Chem.* **2006**, 27, 1787-1799.

[57]. Bethe, H.; Salpeter, E. A Relativistic Equation for Bound-State Problems. *Phys. Rev.* **1951**, 84, 1232-1242.

[58]. Shishkin, M.; Kresse, G. Implementation and performance of the frequency-dependent G W method within the PAW framework. *Phys. Rev. B*, **2006**, 74, 035101-035115.

[59]. Zhang, Z., Liu, L., Fang, W. H., Long, R., Tokina, M. V., & Prezhdo, O. V. (2018). Plasmon-mediated electron injection from Au nanorods into MoS2: traditional versus photoexcitation mechanism. *Chem* **2018**, 4, 1112-1127.

**TOC GRAPHICS**

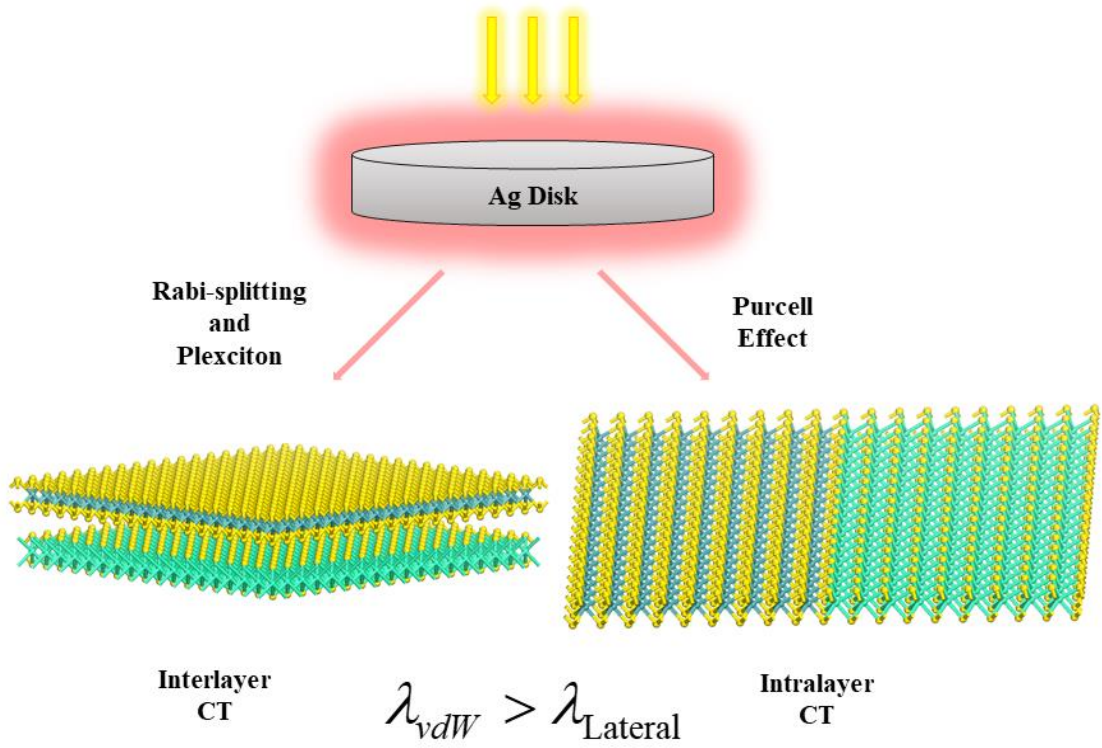